\documentclass[aps,twocolumn]{revtex4}
\usepackage{graphicx}
\usepackage{epsfig}
\begin{document}
\title{High Temperature Universal Properties of Atomic Gases Near 
Feshbach Resonance with Non-Zero Orbital Angular Momentum}
\author{Tin-Lun Ho and Neldeltcho Zahariev}
\address{Department of Physics, The Ohio State University,
Columbus, Ohio 43210}

\begin{abstract}
We show that the high temperature behavior of atomic gases near Feshbach Resonance with non-zero orbital angular momentum
$(\ell >0)$ belong to a universality class different from that of $s$-wave resonances. The universal interaction energy is $2(2\ell +1)$ times larger than that of the $s$-wave when approaching the resonance from the atomic side, but is essentially zero on the molecular side; 
contrary to $s$-wave resonances where interaction energies on both sides are the same except for a sign change. The measurement of these universality properties should be feasible in current experiments. 
\end{abstract}
\maketitle

In the last eighteen months, Feshbach resonance has been used with great success to achieve molecular condensates\cite{MBEC} and fermion superfluids\cite{FC}. At the same time, the physics of Feshbach resonance poses a challenging many-body problem due to its non-perturbative nature. Through Feshbach resonance, a pair of atoms can be converted into molecules. Just before and after a bound pair is formed, the scattering length $a_{s}$ diverges, changing from positive infinity to negative infinity across the resonance. These divergences render the usual perturbative scheme in terms of the gas parameter $n^{1/3} a_{s}$ inapplicable (where $n$ is the density), and is the source of difficulty in theoretical treatment. At the same time, it implies that the system can exhibit ``universal behavior", provided there are no other anomalously large length scales in the system. The reason is that the diverging scattering length must disappear from the physical properties of the system. If all other length scales are smaller than the inter-particle spacing $n^{-1/3}$ and the thermal wavelength $\lambda$, the thermodynamic functions of the system can only depend on these lengths and not on any microscopic properties, and is in this sense universal. Such universal behavior has indeed been observed in many recent experiments\cite{Duke,ENSenergy}.

At present, most experiments are on $s$-wave Feshbach resonance. However, there are many other resonances with non-zero orbital angular momenta $(\ell> 0)$. Very recently, Salomon's group at ENS\cite{ENSp} has reported a reversible production of molecules in Fermi gas of $^{6}$Li across a $p$-wave resonance, paving the way for condensation of $p$-wave molecules and realization of $p$-wave fermion superfluid in the future. There is also a very exhaustive recent study of the Feshbach resonances of the Bose gas $^{133}$Cs\cite{Vuletic}. Resonances upto $\ell=4$ have been observed. Many of the molecules appear to have long lifetimes\cite{Vuletic}. It is natural to ask how quantum gases near $\ell >0$ resonances differ from those of $s$-wave resonances. Will there be universal behavior near resonance? And if there is, do they belong to the same class as $s$-resonances. What is the nature of the $\ell>0$ fermion superfluids in the strongly interacting region, and what is the nature of the molecular condensates? In this paper, we shall address the issue of universality near $\ell > 0$ resonance. The studies of the $\ell>0$ fermion superfluids and molecular condensates are quite involved and will be discussed separately. 
 
Since the physics at resonance is non-perturbative, it is useful to have exact results that can be used as a guide for theoretical approximations. Such exact results are possible in the high temperature regime, where the grand potential can be expanded in powers of the fugacity\cite{HM}. Note that the conceptual problem related to the diverging scattering length does not disappear at high temperatures. For example, it is well known that the interaction energy density for $s$-wave scattering far from resonance is $gn^2$ at all temperatures, where $g=4\pi \hbar^2 a_{s}/M$, and $M$ is the mass of the atom. This expression can not persist at resonance because of diverging $g$. In the case of $s$-wave, one of us has shown recently that the interaction energy density at high temperatures will change from $gn^2$ to the universal value $(3nk_{B}T/2)(n\lambda^3/2^{3/2})$ as one approaches the resonance\cite{HM}. Here, we shall perform similar exact calculations for $\ell >0$ resonance. We have in mind a two component Fermi gas with identical number in each component, $n_{\uparrow}=n_{\downarrow}=n/2$. Our exact results should be useful for future experiments.

Our calculation reveals several remarkable features: 
{\bf (A)} At high temperatures or low densities, the interaction energy densities $\epsilon_{\rm int}^{}$ for {\em all} $\ell>0$ resonances have identical behavior across resonance, apart from a trivial degeneracy factor $(2\ell+1)$. They are, however, different from that of $s$-wave. {\bf (B)} If we denote $\epsilon_{\rm int}^{(a)}$ and $\epsilon_{\rm int}^{(m)}$ the interaction energy density when approaching the resonance from the atomic or molecular side, (defined as the side of resonance where bound state is absent or present respectively), then for $\ell =0$, $\epsilon_{\rm int}^{(a)}$ and $\epsilon_{\rm int}^{(m)}$ are antisymmetric about the resonance. Such symmetry is lost for all $\ell>0$ resonances. There, $\epsilon^{(m)}_{\rm int}$ is essentialy zero, whereas $\epsilon^{(a)}_{\rm int}$ is large and negative {\em close} to resonance, though it decreases rapidly away from resonance. 
{\bf (C)}  For $s$-wave scattering, the second virial coefficient $b_{2}$ and hence $\epsilon_{\rm int}^{(0)}$ reach universal values $1/2$ and $-(3nk_{B}T/2)(n\lambda^3/2^{3/2})$ respectively at resonance. For $\ell>0$, $b_{2}$ and $\epsilon_{\rm int}^{(a)}$ are twice as big, in addition to a degeneracy factor $(2\ell + 1)$. These differences are due to the presence of the centrifugal barrier in the $\ell >0$ scattering channels. These results are derived below.

{\bf (I) Second virial coefficient and interaction energy density:} Let us recall the well known high temperature expansion of the grand partition function ${\cal Z} = {\rm Tr}e^{-(H-\mu N)/k_{B}T}$ in powers of the fugacity $z=e^{\mu/k_{B}T}$. It is shown by Beth and Ulenbeck in 1937\cite{Ulen} that 
${\cal Z} = {\cal Z} ^{(o)} + 2\sqrt{2} \left( \frac{ V z^2}{\lambda^3} \right) b_{2} + O(z^3)+ ... $
where $``o"$ denotes quantities for the non-interacting system, $V$ is the volume of the system, and $b_{2}$ is the second virial coefficient defined as 
$b_{2}= \sum_{\nu}\left( e^{- \beta E_{\nu}^{(2)}} - e^{- \beta [E_{\nu}^{(2)}]^{(o)}} \right)$, 
where $E_{\nu}^{(2)}$ and $ [E_{\nu}^{(2)}]^{(o)}$ are the energy eigenvalues of a two-particle system with and without interaction.
Separating the bound states from scattering states, Beth and Ulenbeck showed that $b_{2}$ can be expressed as\cite{Ulen}

\begin{figure}[t]
\epsfxsize=2.3in
\epsffile{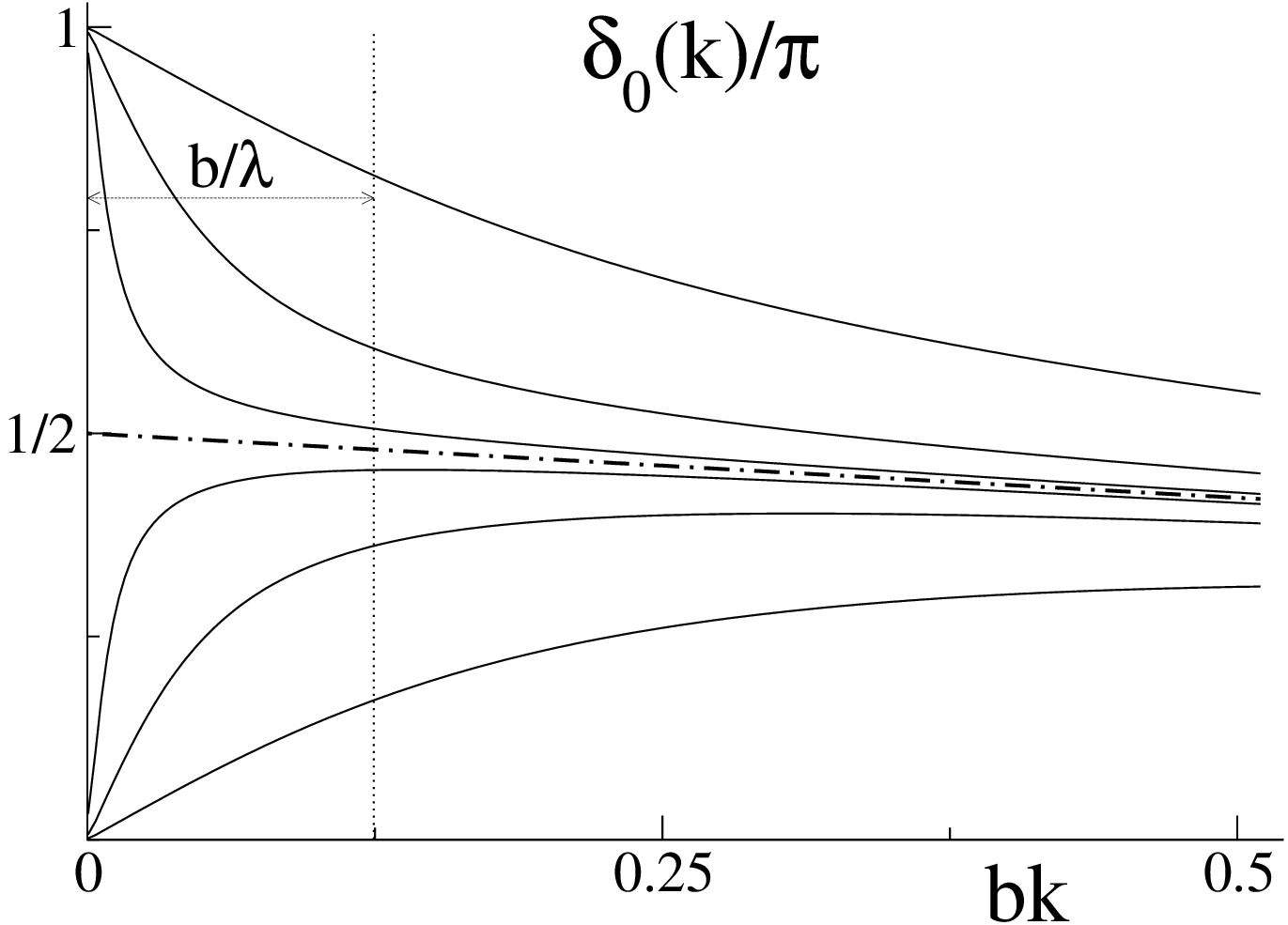}
\caption{$s$-wave phase shifts for a square well potential with width $b$ and various depths. Different curves starting from bottom to top correspond to $a_{0}/b$ $ = -5$, $-20$, $ -100$, $\pm \infty$ (dotted-dash line), $100$, $20$, $5$. The effective range at resonance is $r_{0}=b$. A typical thermal wave length within the universal range of temperatures is also indicated.}
\end{figure}
\begin{equation}
b_{2} = \sum_{b.s.} e^{\frac{|E_b|}{k_BT}}+\sum_{\ell=0}^{\infty}(2\ell+\!1)\int_0^\infty\frac{{\rm d}k}\pi\frac{{\rm d}\delta(k)}{{\rm d}k}e^{-\frac{\lambda^2k^2}{2\pi}},
\label{b2} 
\end{equation}
where $E_{b}$ is the energy of the two-body bound state, $\delta_{\ell}(k)$ is the phase shift of the $\ell$-th partial wave, and $\lambda=\sqrt{ 2\pi \hbar^2/(Mk_{B}T)}$ is the thermal wavelength. From the fugacity expansion, it is straightforward to show \cite{HM} that the energy density can be expanded in the small parameter $n \lambda^3$, and $\epsilon(T, n) = \epsilon_{\rm kin}(n,T)+ \epsilon_{\rm int}(n,T)$, where $\epsilon_{\rm kin}(n,T) = \frac{3nk_{B} T}{2} \left( 1 \pm \frac{ n \lambda^3}{2^{7/2}} + ... \right)$, 
is the kinetic energy of an ideal gas, and
\begin{equation}
\epsilon_{\rm int} = \frac{3k_{B} T n }{2} \left( n \lambda^3 \right)
\left[ - \frac{b_{2}}{\sqrt{2}} + \frac{\sqrt{2}}{3} T \frac{\partial b_{2}}{\partial T} \right], 
\label{int} \end{equation}
is the interaction energy density. 
The $n\lambda^3$ term in $\epsilon_{\rm kin}$ is the well known statistical correction, with $+$ and $-$ sign for bosons and fermions respectively. 

{\bf (II) Effect of centrifugal barrier on the binding energy and the phase shifts:} To understand the general features of the interaction energy near resonance, let us recall the relation between scattering amplitude $f(k,\theta)$ and the phase shift $\delta_{\ell}(k)$, where $\theta$ is the angle between incident and scattered waves. Resolving into partial waves, $f(k,\theta) = \sum_{\ell}(2\ell +1)P_{\ell}(\theta) f_{\ell}(k)$, where $P_{\ell}(\theta)$ are the Legendre polynomials, it is well known that $f_{\ell}(k)$ is related to the phase shift $\delta(k)$ of the $\ell$-th partial wave as 
$f_{\ell}(k) = k^{-1} (\cot\delta_{\ell} -i)^{-1}$. 
For $kb<\!<1$, where $b$ is the range of the scattering potential, $\delta_{\ell}(k)$ has an expansion\cite{LL} 
\begin{equation}
\cot\delta_{\ell}(k) = \frac{(kb)^{-2\ell}}{k} \left( - \frac{1}{a_{\ell}} + \frac{r_{\ell}k^2}{2}+ ...\right),\,\,\,
kb<\!<1,
\label{delta} \end{equation}
where $a_{\ell}$ and $r_{\ell}$ are the scattering length and effective range of the $\ell$-th partial wave 
respectively. The low energy scattering amplitude can then be written as\cite{LL2}
\begin{equation}
f_{\ell}(k) = \frac{ (kb)^{2\ell} }{ -1/a_{\ell} + r_{\ell} k^2/2 - ik (kb)^{2\ell} }, \,\,\,\,\,
kb<\!<1, 
\label{fell} \end{equation}
It is also well known that by analytically continuing $k$ in the complex plane, if $f_{\ell}(k)$ has a singularity on the positive imaginary axis, $k=i\kappa$, $\kappa>0$, then the system has a bound state with energy $E_{b} = - \hbar^2 \kappa^2/M$. 

\begin{figure}[t]
\epsfxsize=2.2in
\epsffile{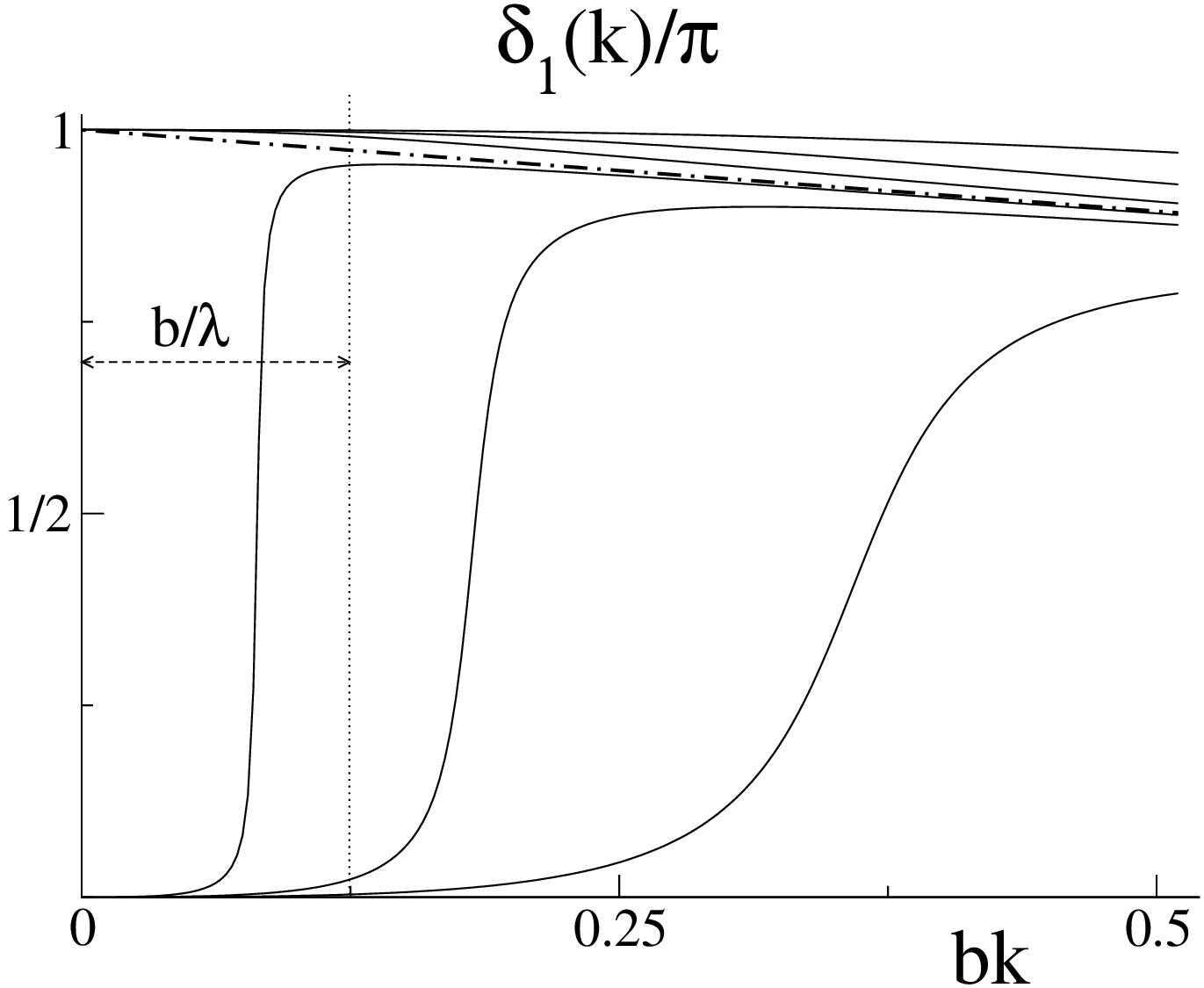}
\caption{Similar to Fig. 1, only for the $p$-wave phase shifts. The successive curves starting from bottom to top are $a_{1}/b$ $ = -5$, $-20$, $-100$, $\pm \infty$ (dotted-dash line), $20$, $5$, $1$. For the square well potential the $p$-wave effective range at resonance is $r_{1}=-3b$.}
\end{figure}

The factor $(kb)^{2\ell}$ in eq.(\ref{fell}) is due to the presence of a centrifugal barrier. It has dramatic effects on the bound state energy and phase shift that appear in eq.(\ref{b2}). For $s$-wave scattering, as $k\rightarrow 0$, eq.(\ref{fell}) becomes $f_{0}(k) \rightarrow 1/(-a^{-1}_{0} - ik)$. The dominant $k$ dependence is the imaginary part in the denominator. When $k$ is analytically continued to the pure imaginary axis, 
 $k= i\kappa$, with $\kappa >0$, $f_{0}(k)$ has a singularity (and hence a bound state) only when $a_{0}>0$; in which case $\kappa = 1/a_{0}$ and $E_{b} = - \hbar^2/(Ma_{0}^2)$. 
In contrast, for $\ell>0$, the low energy behavior of $f_{\ell}$ in eq.(\ref{fell}) is dominated by the real part of the denominator. A bound state occurs when $k^{2} = -\kappa^{2}$ with $\kappa = \sqrt{2/(-a_{\ell} r_{\ell})}$, which is possible only when $a_{\ell}r_{\ell}<0$. The binding energy is $E_{b} = 2\hbar^2/ (Ma_{\ell} r_{\ell})$. Unlike the $s$-wave case, $E_{b}$ is {\em linear} in $1/a_{\ell}$. In addition, it depends on effective range $r_{\ell}$ which is typically of atomic scale. 

Turning to the phase shifts, eq.(\ref{delta}) implies that 
\begin{equation}
\frac{{\rm d}\delta_{\ell}(k)}{{\rm d} k} = \frac{ (kb)^{2\ell}\left[ - (2\ell + 1)a_{\ell}^{-1} + 
(2\ell -1) r_{\ell}k^2/2 \right] }{
 \left[ -a_{\ell}^{-1} + r_{\ell} k^2/2\right]^2 + (kb)^{4\ell}k^2 }
\label{dddk} \end{equation}
When $\ell=0$, the denominator is $k^2 + (a^{-1}_{0} - r_{0}k^2/2)^2$. Near resonance $|a_{0}|>>r_{0}$ and one can ignore $r_{0}$ to second order in $k$, and hence 
\begin{equation}
\frac{{\rm d}\delta_{0}(k)}{{\rm d} k} = \frac{-a_{0}}{1 + a_{0}^2 k^2 } + O\left(r_{0} k^{2}\right), 
\label{dds} \end{equation}
which is a half Lorenzian in the range of positive $k$. Note that this result is valid on both sides (molecular and atomic) of the resonance. In contrast, for $\ell>0$, the term $(kb)^{2\ell}$ is negligible in the 
denominator of eq.(\ref{dddk}), so the effective range $r_{\ell}$ can not be ignored. This makes ${\rm d}{\delta_{\ell}}/{\rm d} k$ very different on the two 
sides of the resonance. On the atomic side where bound states are absent, we have $a_{\ell}r_{\ell}>0$. Near $k=k_{c} \equiv \sqrt{2/(a_{\ell}r_{\ell})}$, we have 
\begin{equation}
\frac{{\rm d}\delta_{\ell}(k)}{{\rm d} k} =
\frac{\Gamma_{\ell} (k_{c})}{ (k-k_{c})^2 + \Gamma_{\ell}(k_{c})^2} , \,\,\,\,\,\, a_{\ell}r_{\ell}>0, 
\label{ddell} \end{equation}
where $\Gamma_\ell(k) = |(kb)^{2\ell}/r_{\ell}|$. Eq.(\ref{ddell}) is a very narrow Lozenzian because $\Gamma_{\ell}(k_{c})/k_{c} = |(2b/a_{\ell})^{\ell-1/2} (b/r_{\ell})^{\ell+1/2}|<\!<1$. 
Thus, unlike the $s$-wave case where ${\rm d} \delta_{0}(k)/{\rm d}k$ is a half Lorenzian in the positive $k$-axis, ${\rm d} \delta_{\ell}(k)/{\rm d}k$ is a {\em full} Lorenzian in the same range of $k$ for $\ell >0$. 
On the molecular side where $a_{\ell}r_{\ell}<0$, 
${\rm d}\delta(k)/{\rm d}k$ has no resonance structure. Since it is proportional to $\propto (kb)^{2\ell}$, 
$\delta_{\ell}$ is exceedingly small. 

\begin{figure}[t]
\epsfxsize=2.5in
\epsffile{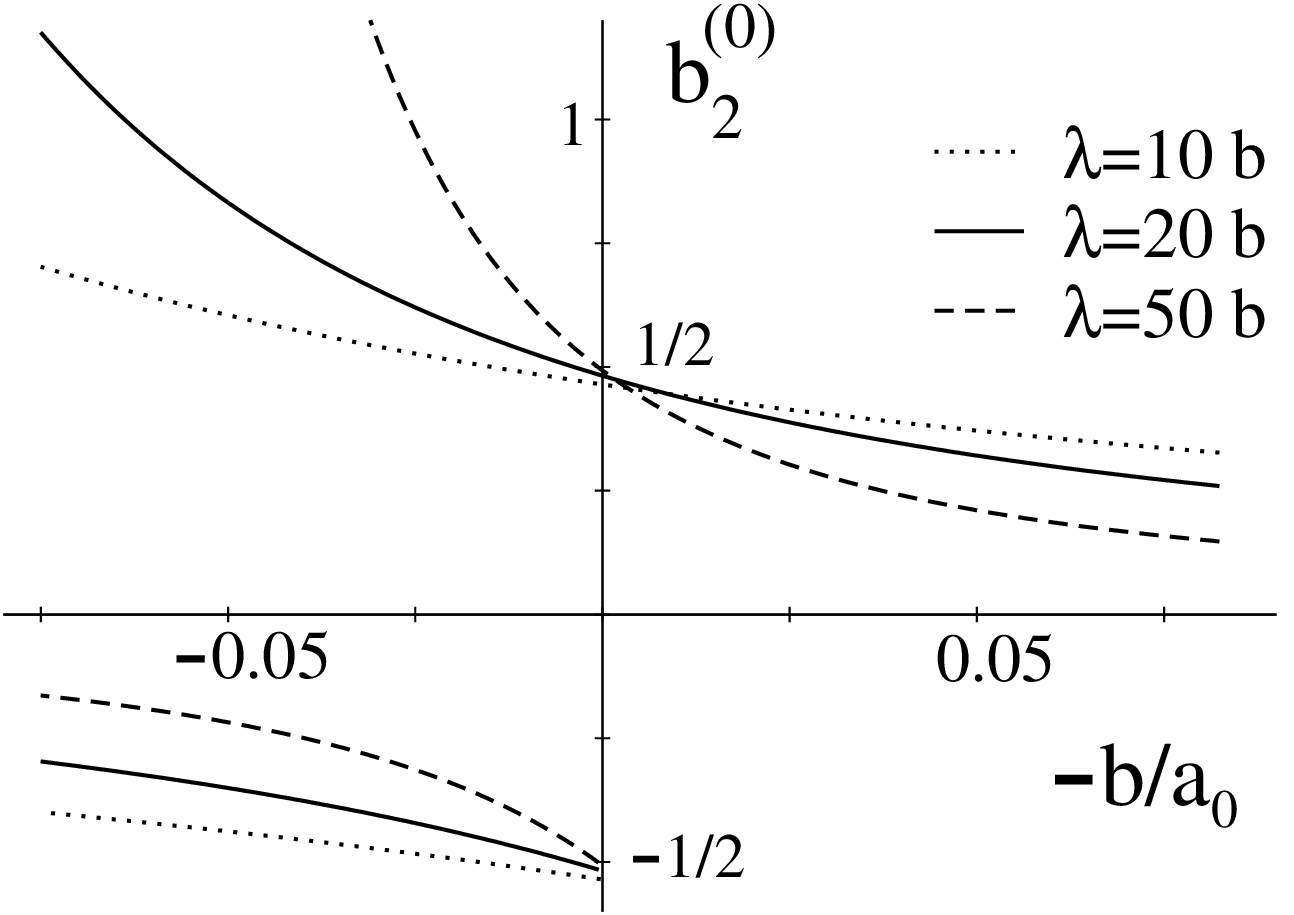}
\caption{$b_{2}$ for $\ell\!=\!0$. For $^6$Li and $b=10{\rm nm}\approx 200 a_B$ the corresponding temperatures  are dashed, solid and dotted lines $50\mu$K, $12\mu$K and $2\mu$K. The deviation of $b_{2}$ from $\pm 1/2$ at resonance is due to effective range corrections.}
\end{figure}

The differences between $\delta_{0}$ and $\delta_{\ell}$ are shown explicitly in Fig. 1 and Fig. 2 (for $\ell=1$). Let us consider both cases very close to resonance on the atomic side. For $\ell=0$, $\delta_{0}$ rises quickly from 0 to $\pi/2$ as $k$ increases. It stays close to $\pi/2$ within the region $a_{0}^{-1}<\!< k <\!< r_{0}^{-1}$, where $\cot\delta_{0}$ satisfies $0< \cot\delta_{0}<\!<1$ (see eq.(\ref{delta})). It then begins to decrease as $k$ increases. (Note that all $\delta_{\ell}(k)$ must vanish as $k\rightarrow \infty$, since the effect of scattering potential becomes unimportant at very high energies). For $\ell>0$, however, 
$\delta_{\ell}(k)$ remains close to zero until $k$ approaches $k_{c}=\sqrt{2/(a_{\ell}r_{\ell})}$. It rises quickly from $0$ to $\pi$ as $k$ passes through $k_{c}$. It stays close to $\pi - 0^{+}$ within the region $\sqrt{2/(a_{\ell}r_{\ell})}<\!<k <\!<
b^{-1}|r_{\ell}/b|^{1/(2\ell -1)}$ where $\cot\delta_{\ell}<\!< -1$, and then decreases as $k$ increases. 
These marked difference in behavior lead to the differences in ${\rm d}\delta_{\ell}/{\rm d} k$ shown in eq.(\ref{dds}) and (\ref{ddell}).

\begin{figure}[t]
\epsfxsize=2.5in
\epsffile{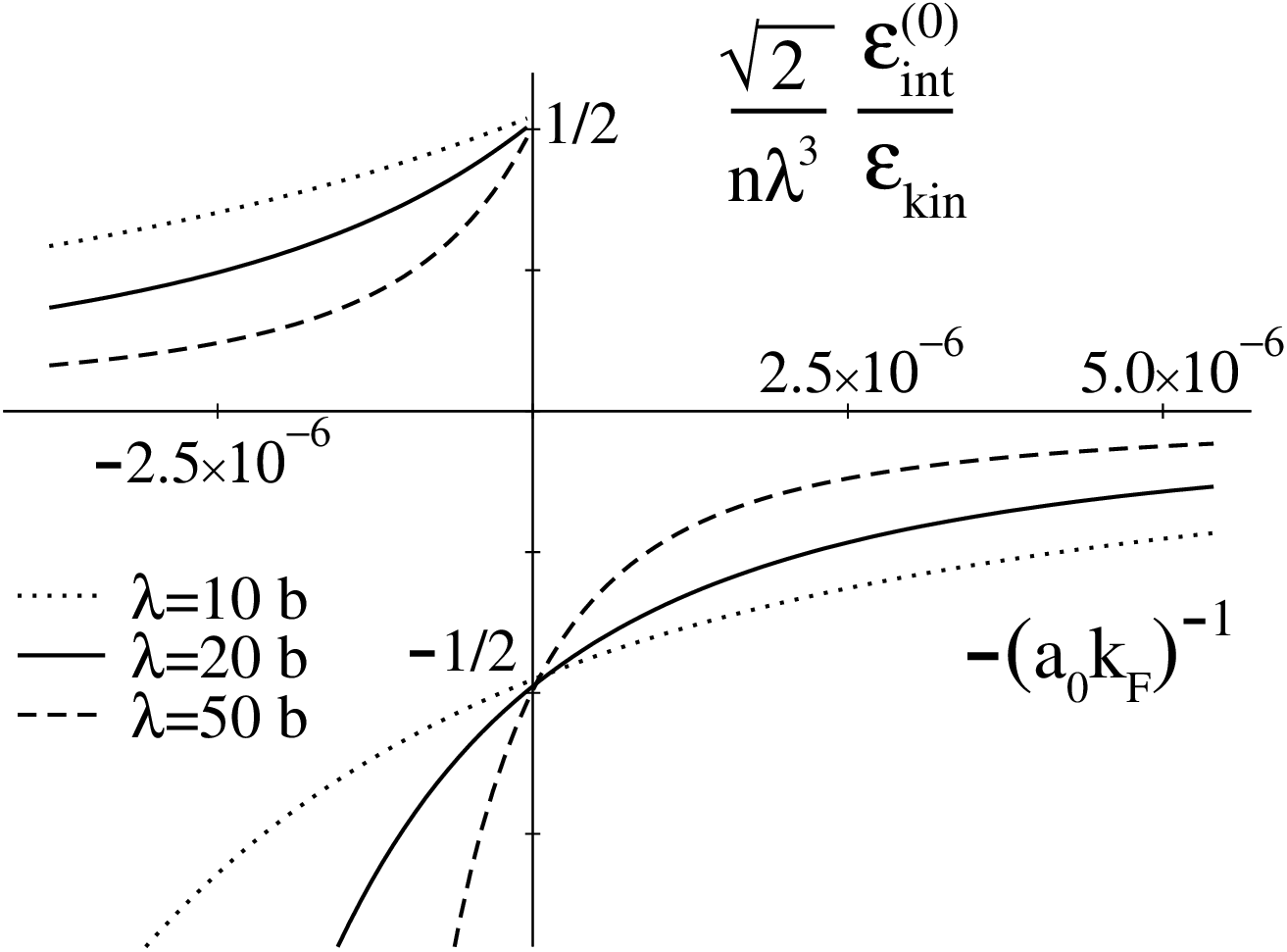}
\caption{$\epsilon_{int}$ for $\ell=0$. The same temperatures as in Fig. 3. $k_{F}$ is related to density as $k_{F}^3=3\pi^2 n$, which we choose to be $10^{12}{\rm cm}^{-3}$.}
\end{figure}

{\bf (III) Final Results:} Let us first explain the reason for the emergence of universal behavior near resonance before deriving the accurate formulas. 
The first term in eq.(\ref{b2}) is certainly universal near resonance since $E_{b}\rightarrow 0$ as the bound state emerges. 
The second term in eq.(\ref{b2}) is the contribution due to scattering states. 
This integral is cut off by $1/\lambda$, and is roughly given by 
\begin{equation}
b_{2}^{\!(\ell)}\! \approx\! (2\ell\! +\!1)\!\! \int_{0}^{\lambda^{\!-1}}\!\! \frac{ {\rm d} k}{\pi} \frac{ {\rm d} \delta_{\ell}(k)}{{\rm d} k} \!=\! \frac{(2\ell\!\!+\!\!1)}{\pi}\! \left[\delta_{\ell}(\lambda^{\!-1}) \!-\! \delta_{\ell}(0)\right]
\label{approx} \end{equation}
Near resonance, the $s$-wave phase shift $\delta_{0}(k)$ quickly rises from zero to $\pi/2$ and stays basically flat within the region $a^{-1/2}_{0}<\!<k<\!<r_{0}^{-1}$, where $0< \cot \delta_{0} <\!< 1$. (See eq.(\ref{delta}) and also Fig. 1). 
We then have 
\begin{equation}
b_{2}^{(0)} \approx 1/2 \,\,\,\,\,\,\, {\rm for}\,\, \,\,\,\, r_{o}<\!<\lambda <\!< n^{-1/3}. 
\label{appb20} \end{equation}
The condition $\lambda n^{1/3}<\!<1$ is automatically enforced because of the low fugacity\cite{HM}. Eq.(\ref{appb20}) shows that universal behavior {\em only} emerges when the temperature is sufficiently high to achieve low fugacity limit but sufficiently low so that the thermal wavelength is still much larger than the effective range, which is typically the range of the potential. 
In fact, at very high temperatures where $\lambda\rightarrow 0$, $b_{2}$ vanishes for all $\ell$ since $\delta(\lambda^{-1})\rightarrow \delta(\infty)=0$. However, such high temperature range is not of interests in current experiments. 

For $\ell>0$, $\delta_{\ell}(k)$ rises rapidly from 0 to $\pi$ as $k$ passes through $k_{c}=\sqrt{2/(a_{\ell}r_{\ell}})$. It stays close to $\pi$
within the region $\sqrt{2/a_{\ell}r_{\ell}} <\!<k <\!< b^{-1} (r_{\ell}/b)^{1/(2\ell -1)}$, where $\cot\delta_{\ell} <\!< -1$. (See also fiure 2). This means that 
 \begin{equation}
b_{2} \approx (2\ell +1) \,\,\,\,\, {\rm for} \,\,\,\,\, b(b/r_{\ell})^{1/(2\ell -1)} <\!<\lambda <\!< n^{-1/3}. 
\label{appb2ell} \end{equation}

Having explained the emergence of universality near resonance, we now derive the precise formula for interaction energy for all region of scattering length. The case of $s$-wave resonance has already been discussed in ref.\cite{HM}. Substituting eq.(\ref{dds}) into eq.(\ref{b2}), we have for $\ell=0$
 \begin{equation}
 b_{2} = \sum_{b} e^{\frac{|E_{b}|}{k_{B}T}} - \frac{ {\rm sgn}(a_{0})}{2} \left[ 1 - {\rm erf}(x)\right] e^{x^2} + O\left( \frac{r_{0}}{\lambda}\right), 
\label{b2a0final}\end{equation}
where ${\rm erf}(x)$ is the error function and $x=\lambda/\sqrt{2\pi}a_{0}$. 
The behaviors of $b_{2}$ and $\epsilon_{\rm int}$ calculated from eq.(\ref{b2}) and (\ref{int}) are shown in figures 3 and 4. The results in these figures are exact
calculations for the square well potential and are found to be indistinguishable from eq.(\ref{b2a0final}) when effective range correction is included. 
The antisymmetry of $b_{2}$ and $\epsilon_{\rm int}$ on different sides of the resonance is due to the antisymmetry of eq.(\ref{dds}) in $1/a_{0}$. From eq.(\ref{b2}) and 
(\ref{dddk}), one can easily show that at resonance, $b_{2} = \pm 1/2 - 2^{3/2}r_{0}/\lambda$, which confirms 
eq.(\ref{appb20}). The corresponding universal value for the energy density is 
 $\epsilon_{\rm int} = (3nk_{B}T/2)(n\lambda^3/2^{3/2}) + O(r_{0}/\lambda)$. 
 
\begin{figure}[t]
\epsfxsize=2.7in
\epsffile{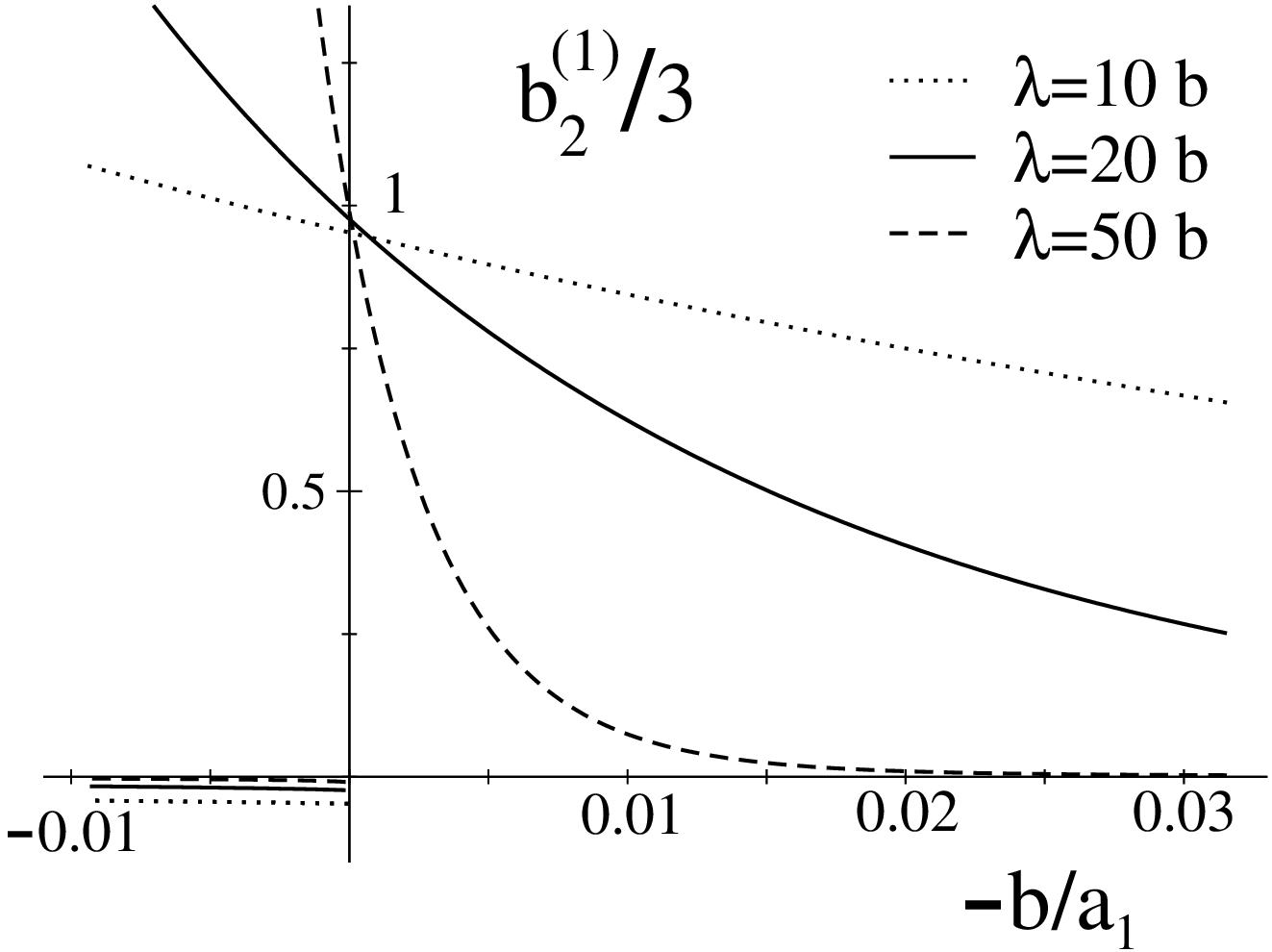}
\caption{$b_{2}$ for $\ell=1$. The dashed, solid, and dotted lines corresponds to $T= 50, 12, 2 \mu$K. The deviations of $b_{2}$ from $1$ on the atomic side and $0$ on the molecular are due to effective range corrections.}
\end{figure}
 
\begin{figure}[t]
\epsfxsize=2.7in
\epsffile{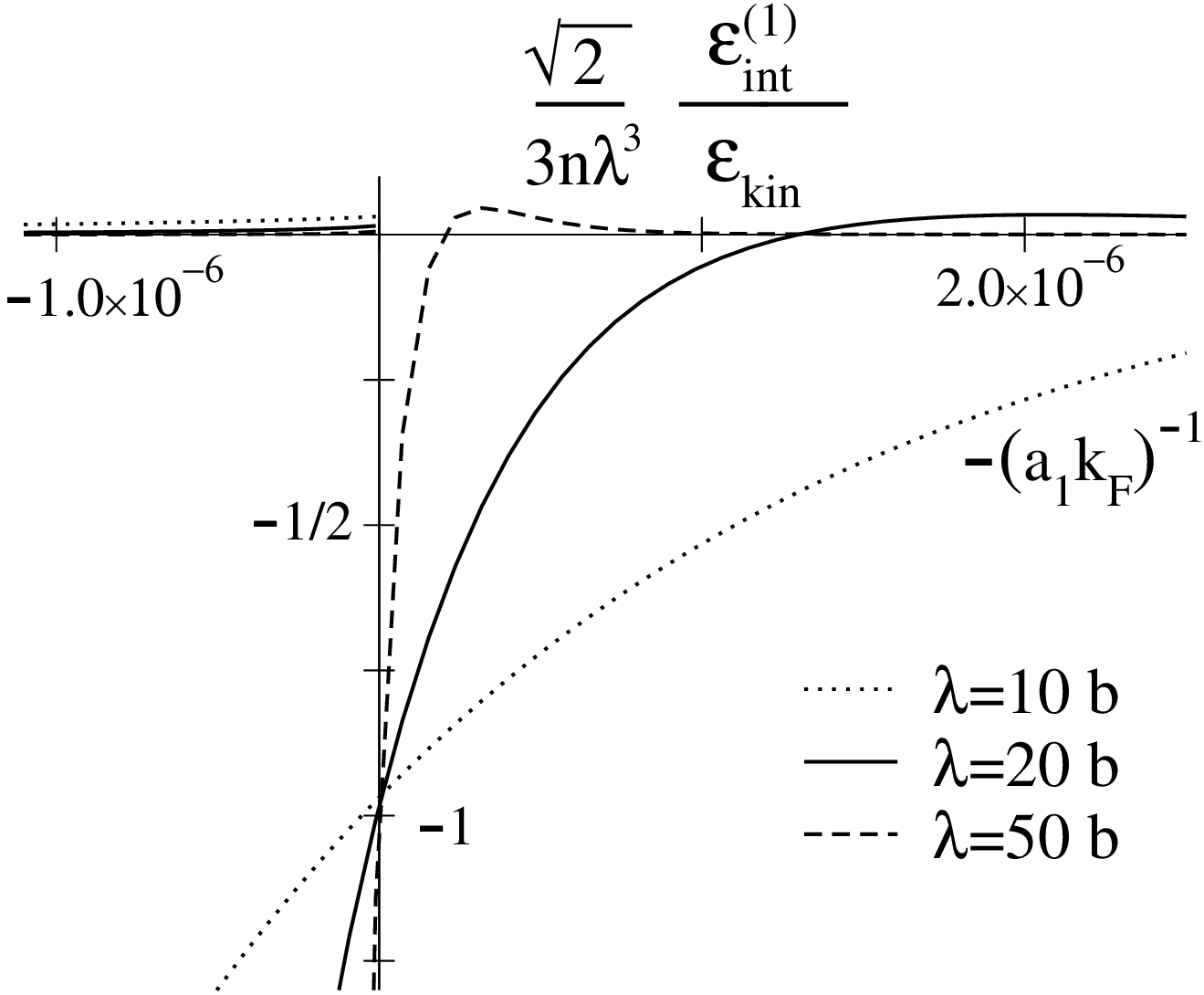}
\caption{$\epsilon_{\rm int}$ for $\ell=1$. Same parameters as in Fig. 4. Note that the molecular branch (top left curves) $\epsilon_{\rm int}^m$ is close to $0$ and the atomic branch at resonance is larger than that of the $s$-wave.}
\end{figure}

For $\ell>0$, eq.(\ref{ddell}) and (\ref{b2}) imply that on the atomic side 
\begin{equation}
b_{2}^{} = (2\ell\!+\!1)e^{-\frac{2\lambda}{\pi a_{\ell}r_{\ell}}} 
{\rm erfc}\left( \frac{\lambda}{\sqrt{2 \pi } r_{\ell}} 
 \left( \frac{ 2b^2}{ a_{\ell} r_{\ell} } \right)^{\ell} \right) + (... ),
\label{b2ellfinal}\end{equation}
where ${\rm erfc}(x) = 1 - {\rm erf}(x)$, and $(...)$ means correction of the order of 
$O\left( \frac{b}{r_{\ell}} \left[ \frac{b}{\lambda}\right]^{2\ell -1} \right) $. The behaviors of $b_{2}$
and $\epsilon_{\rm int}$ for $\ell=1$ are shown in figures 5 and 6. They are exact calculations for a square well and are in excellent agreement with eq.(\ref{b2ellfinal}). Approaching the resonance from the atomic side, we see from eq.(\ref{dddk}) that 
$b_2 =(2\ell +1)\left( 1 - (2\ell -1)!! \pi^{\ell} 2^{3/2} \frac{b}{r_{\ell}} \left[ \frac{b}{\lambda}\right]^{2\ell -1}\right)$, which verifies eq.(\ref{appb2ell}). The interaction energy density is 
$\epsilon_{\rm int} = 2(2\ell +1) (3nk_{B}T/2)(n\lambda^3/2^{3/2})$
$+ O\left( \frac{b}{r_{\ell}} \left[ \frac{b}{\lambda}\right]^{2\ell -1} \right) $. 

On the molecular side, $\epsilon_{\rm int}$ is zero to the zeroth order in $r_{0}/\lambda$ because of the exceedingly small phase shifts $(\sim k^{2\ell +1})$.
Far from resonance, $\delta_{\ell}(k) = -a_{\ell}k (kb)^{2\ell}$, we have for all $\ell$ and on both sides of the resonance, $\epsilon_{\rm int}^{(\ell)} = g_{\ell} n_{\uparrow}n_{\downarrow} \left(\frac{b}{\lambda}\right)^{2\ell}
\pi^{\ell} (2\ell +1)!! 
 \left( [1-\ell] - \frac{ \pi}{2} \frac{a_{\ell}r_{\ell}}{\lambda^2} \ell (2\ell + 3) \right)$,
where $g_{\ell}= 4\pi\hbar^2 a_{\ell}/M$, and $n_{\uparrow}=n_{\downarrow}=n/2$. This shows that away from resonance, the interaction energy in the $\ell>0$ channels can not be described by a single length scale $a_{\ell}$ as in the case of $s$-wave scattering. 
We have thus established the results (A) to (C) in the Introduction. 

This work is supported by NASA GRANT-NAG8-1765 and NSF Grant DMR-0109255.

\end{document}